\documentclass[conference]{IEEEtran}
\IEEEoverridecommandlockouts
\ifCLASSOPTIONcompsoc
  
  \usepackage[nocompress]{cite}
\else
  \usepackage{cite}
\fi

\usepackage{cite}
\usepackage{amsmath}
\usepackage{amssymb}
\usepackage{amsfonts}
\usepackage{graphicx}
\usepackage{textcomp}
\usepackage{xcolor}
\usepackage{multirow}
\usepackage{soul}
\usepackage{tikz}
\usepackage{algorithm}
\usepackage{algpseudocode}
\usepackage{colortbl}
\usepackage{bm}
\usepackage{float}
\usepackage{filecontents}
\usepackage{CJK}
\usepackage{algorithm}
\usepackage{algorithmicx}
\newcommand\encircle[1]{%
\tikz[baseline=(X.base)] 
  \node (X) [draw, scale=0.75, shape=circle, inner sep=0, fill=black, text=white, minimum size=0em] {\strut #1};}

\newcommand{\cmmnt}[1]{}  

\def\BibTeX{{\rm B\kern-.05em{\sc i\kern-.025em b}\kern-.08em
    T\kern-.1667em\lower.7ex\hbox{E}\kern-.125emX}}

\ifCLASSINFOpdf
\else
\fi

\usepackage{todonotes}

\usepackage[a4paper, total={184mm,239mm}]{geometry}
\def\BibTeX{{\rm B\kern-.05em{\sc i\kern-.025em b}\kern-.08em
    T\kern-.1667em\lower.7ex\hbox{E}\kern-.125emX}}

\begin{document}
\newcommand{\algrule}[1][.2pt]{\par\vskip.5\baselineskip\hrule height #1\par\vskip.5\baselineskip}

\title{\huge OISA: Architecting an Optical In-Sensor Accelerator for Efficient Visual Computing \vspace{-0.5em}
}

\author{Mehrdad Morsali$^\dagger$, Sepehr Tabrizchi$^{\ddagger}$, Deniz Najafi$^\dagger$, Mohsen Imani$^\S$,\\ Mahdi Nikdast$^*$,   Arman Roohi$^{\ddagger}$, and Shaahin Angizi$^\dagger$ \vspace{0.3em}\\ \small $^\dagger$Department of Electrical and Computer Engineering, New Jersey Institute of Technology, Newark, NJ, USA\\
$^\ddagger$School of Computing, University of Nebraska–Lincoln, Lincoln, NE, USA\\
$^\S$Department of Computer Science, University of California Irvine, Irvine, CA, USA\\
$^*$Department of Electrical and Computer Engineering, Colorado State University, Fort Collins, CO, USA\\ 
m.imani@uci.edu, mahdi.nikdast@colostate.edu, aroohi@unl.edu, shaahin.angizi@njit.edu \vspace{-2em}
\\}

\maketitle
\begin{abstract}
Targeting vision applications at the edge, in this work, we systematically explore and propose a high-performance and energy-efficient \ul{O}ptical \ul{I}n-\ul{S}ensor \ul{A}ccelerator architecture called OISA for the first time. Taking advantage of the promising efficiency of photonic devices, the OISA intrinsically implements a coarse-grained convolution operation on the input frames in an innovative minimum-conversion fashion in low-bit-width neural networks. Such a design remarkably reduces the power consumption of data conversion, transmission, and processing in the conventional cloud-centric architecture as well as recently-presented edge accelerators. Our device-to-architecture simulation results on various image data-sets demonstrate acceptable accuracy while OISA achieves 6.68 TOp/s/W efficiency. OISA reduces power consumption by a factor of 7.9 and 18.4 on average compared with existing electronic in-/near-sensor and ASIC accelerators.
\end{abstract}


\section{Introduction}
While the Internet of Things (IoT) has become ubiquitous, it still lacks inherent intelligence and heavily depends on cloud-based decision-making. In such a cloud-centric scenario, a substantial part of data generated by IoT's sensors is left unprocessed or unanalyzed \cite{song2022reconfigurable,xu2020macsen}. Vision sensors typically convert light into electrical signals, which are then saved, processed, transmitted, and utilized. This involves converting all pixels into predetermined digital values with a constant bit depth, such as 8 bits \cite{xu2020macsen,angizi2023pisa}. Reportedly, the majority of power consumption (over 96\% \cite{xu2020macsen,angizi2023pisa}) in traditional vision sensors comes from pixel value conversion and storage. This is primarily due to the memory and compute-intensive computing algorithm and the lack of processing capabilities of current IoT devices restricted by power and area factors \cite{tang2019considerations,angizi2023near}. To tackle these challenges, a shift from a cloud-centric to a thing-centric (data-centric) approach is required, where the IoT node processes the sensed data locally \cite{hsu2019ai}.

Recently, there has been research into developing smarter CMOS image sensors that can accelerate Deep Neural Network (DNN) workloads. One method is to integrate CMOS image sensors and processors on a single chip, referred to as Processing-Near-Sensor (PNS) \cite{morsali2023design,hsu20200,yamazaki20174,angizi2018cmp,angizi2019mrima}. Another approach involves integrating computation units with individual pixels called Processing-In-Sensor (PIS). The PIS platform \cite{xu2020macsen,xu2021senputing,tabrizchi2023appcip,angizi2023pisa} processes pre-Analog-to-Digital Converter (pre-ADC) data before transmitting it to the on/off-chip processor \cite{el1999pixel,song2022reconfigurable}. However, they still suffer from energy-hungry ADC, DAC, and sense amplifiers \cite{xu2020macsen,angizi2023pisa}. Due to the limited resources of PIS, it has not been feasible to deploy all DNN layers into the pixel array. Therefore, most studies have focused on accelerating the first layer in an analog or digital domain and submitting the remaining layers to a digital accelerator. Nevertheless, three significant challenges have yet to be addressed in current PIS/PNS designs
\textit{
(i) Current designs still suffer from energy-hungry peripherals and ADC/DAC units even reduced \cite{choi2015energy,xu2020macsen,hsu2019ai} for sensing and computing; 
(ii) the in-/near-sensor computation imposes a large area overhead and power consumption in more recent PNS/PIS units and typically requires extra memory for intermediate data storage \cite{angizi2023pisa,tabrizchi2023appcip,song2022reconfigurable,roohi2023pipsim}; and 
(iii) the computation speed has been constrained by the electronic systems (operating at a few GHz) that inherently lack the capability to support both high speeds and the extensive parallelism found in optical systems approaching
the photodetection rate ($>$100GHz)\cite{sunny2021robin,sunny2021crosslight,cheng2020silicon}.}

While silicon photonics has already established its efficacy in enabling high-throughput communication and computation across various domains \cite{sunny2021crosslight,sunny2021robin}, in this work, we systematically explore the potential of deploying it on edge devices. We propose an Optical In-Sensor Accelerator architecture named OISA that leverages the energy efficiency and low latency features of photonic devices and minimizes signal conversion in low-bit-width neural networks to eliminate the need for power-hungry ADC and DACs. Our novel contributions to this work are as follows. 
(1) For the first time, we develop an optical in-sensor architecture that has been optimized to efficiently process the 1$^{st}$ layer of DNNs with a centralized kernel-based optical processing unit tuned with microring resonator optical devices, resulting in improved energy-efficiency and speed; 
(2) We design new microarchitectural and circuit-level schemes for OISA supported by novel hardware partitioning and mapping mechanisms; and
(3) We create a bottom-up device-to-architecture evaluation framework and extensively analyze and compare the performance of the proposed designs with prior PIS and ASIC designs.

\section{Background}
\textbf{In-Sensor Accelerators.} 
Boosting throughput and intensifying computation on resource-limited PIS/PNS devices result in elevated temperature, higher power consumption, and increased noise levels. These factors contribute to a decline in accuracy \cite{xu2020macsen,angizi2023pisa,xu2020utilizing}. 
MACSEN \cite{xu2020macsen} as a PIS platform processes the 1$^{st}$-convolutional layer of Binary CNN with the correlated double sampling procedure achieving 1000 FPS speed in computation mode. However, it suffers from humongous area-overhead and power consumption mainly due to the SRAM-based PIS method. In \cite{song2022reconfigurable}, a PIS architecture is designed to support 8-bit activation and weight intended for the first-layer DNN acceleration through pulse modulation. Although the resource utilization of the design improved considerably, the use of power-hungry ADCs in each column, along with capacitors for direct sampling, increased overall energy consumption and area overhead. PISA \cite{angizi2023pisa} enables convolutional operations in the first BNN layer by leveraging non-volatile memory to store network weights. Notably, this architecture reduces the energy consumed on data conversion and transmission. However, the power-demanding write operations in non-volatile memories and the use of ADC for data transfer elevate the overall power consumption of the array. 
In \cite{yamazaki20174}, a PNS architecture was introduced, enhancing resolution while minimizing area overhead. However, PNS faces challenges in addressing the under-utilization problem of the first layer, leading to reduced accuracy. Additionally, the use of ADCs further raises power consumption across the entire array.  AppCiP \cite{tabrizchi2023appcip} implements a folded ADC to decrease comparator count, though the collective power consumption of the ADC units remains an issue. In \cite{lefebvre20217}, the PIS architecture leverages a combination of pixel current and charge-sharing events to reduce the power consumption of ADC to enable feature extraction and region-of-interest detection through current-domain MAC operations. However, the design is limited to row-wise computing rather than performing computations across the entire array.

\begin{figure}[t] 
\centering
\includegraphics [width=0.88\linewidth]{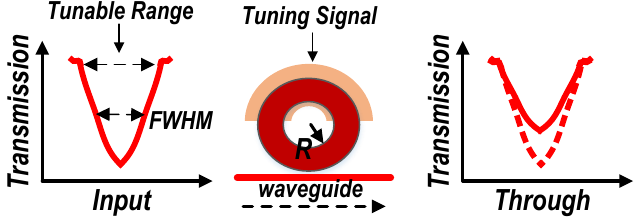}
\vspace{-1.0em}
\caption{MR input and through ports’ spectra after imprinting a parameter onto the signal. A directional coupler transfers the light from the waveguide into the MR to be recombined. This recombination, influenced by the effective refractive index in the MR which is also impacted by the MR's circumference, induces a phase shift in the combined wave. This phase shift leads to interference with the original light's intensity. Tunable range shows the free resonance spectral range of the MR, where FMHW is the full width at half maximum of the resonance spectrum.
}
\vspace{-1.7em}
\label{mr}
\end{figure}
\textbf{Silicon Photonics Accelerators.}
Offering notably elevated operational bandwidth compared to electronic accelerators along with addressing fan-in/fan-out problems make silicon-photonic-based accelerators a promising candidate to accelerate DNN and machine vision applications \cite{sunny2021arxon,sunny2021crosslight}. Such accelerators can be broadly categorized into two primary designs: \textit{coherent} and \textit{non-coherent} architectures. Within the coherent category, a single wavelength is employed for operations, and weight/activation parameters are incorporated into the electrical field amplitude, phase, or polarization of an optical signal \cite{zhao2019hardware}.  Conversely, the noncoherent designs \cite{sunny2021robin,sunny2021crosslight} employ multiple wavelengths each of which capable of conducting computations concurrently. Within coherent architectures, considered in this work, the weight and input parameters are imprinted upon the signal's amplitude. To manipulate individual wavelengths Microring Resonators (i.e., MRs, depicted in Fig. \ref{mr}) can be employed whose central frequency can be actively adjusted (i.e., through tuning mechanisms using, e.g., microheaters or PIN junctions) to selectively interact with specific wavelengths. 
By appropriately tuning the MRs, the incoming light intensity of a specific wavelength can be weighted. In the non-coherent designs \cite{sunny2021crosslight,sunny2021robin}, MRs as a fundamental component store the weight and activation values to be utilized in the MAC operation. During photonic MAC, incoming lights can be multiplied by the value adjusted on the MRs (through applying a tuning signal, see Fig. 1) of the same wavelength.

\begin{figure}[t] 
\centering \vspace{-0.5em}
\includegraphics [width=0.998\linewidth]{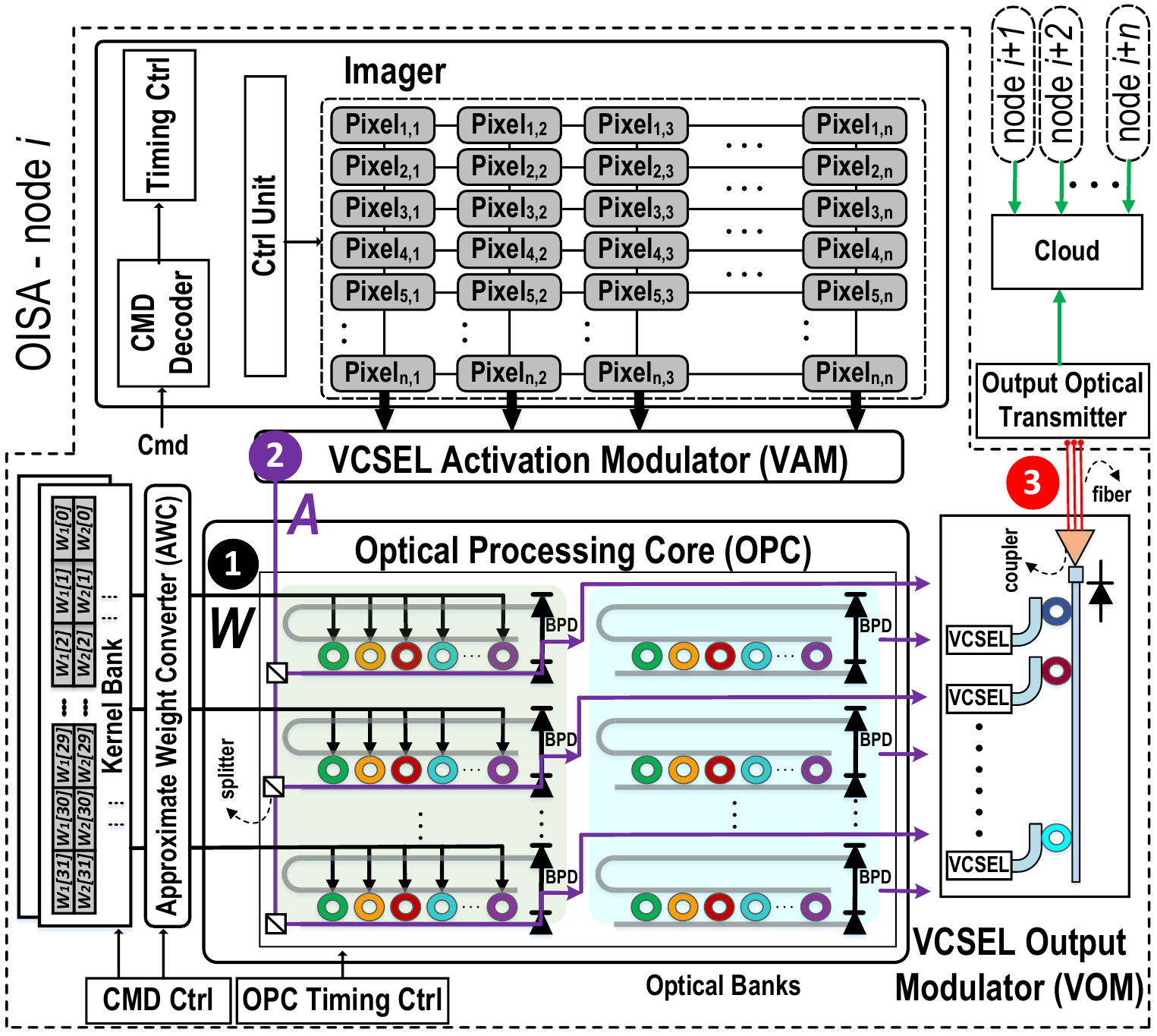}
\vspace{-2.2em}
\caption{OISA architecture with Global shutter CMOS imager, VCSEL-based Activation Modulator (VAM), Optical Processing Core (OPC), and VSCEL-based Output Modulator (VOM).
}
\vspace{-1.7em}
\label{arch}
\end{figure}

\vspace{-0.6em}
\section{OISA Architecture}
We propose OISA as a scalable, high-performance, and low-power solution for real-time image processing at edge devices. 
OISA integrates sensing and processing phases and intrinsically supports a low bit-width (2-bit (Ternary) activation and up to 4-bit weight) MAC operation of the $1^{st}$-layer in Multi-Layer Perceptron (MLPs) and Convolutional Neural Networks (CNNs) while submitting the next layers to an off-chip processor through ultra-fast optical transmitters. The high-level overview of the proposed architecture, denoted by node \textit{i} in a multi-node IoT structure, is shown in Fig. \ref{arch} consisting of \ul{six} key components: 
\textit{(i)} an ADC-less global shutter CMOS imager comprising a n$\times$n conventional pixel structure to capture frames; \textit{(ii)} a Vertical-Cavity Surface-Emitting Lasers (VCSEL)-based Activation Modulator (VAM) that is dedicated to directly modulate every pixel's voltage drop value after exposure to light (activation) with a predetermined wavelength and intensity proportional to the original light that is absorbed by pixel; 
\textit{(iii)} an Approximate Weight  Converter (AWC) to convert weight values stored on kernel banks to a current driving MRs;
\textit{(iv)} an Optical Processing Core (OPC) to execute parallel MAC operation between activation (A) and weight (W) parameters in the photonic domain; 
\textit{(v)} a VSCEL-based Output Modulator (VOM) which is only used during the MLP processing or large kernel processing to break down the MAC operation when the number of elements in partial sum is huge; and
\textit{(vi)} a controller to configure the timing and optical banks to perform data-parallel intra-bank computations. In the following, we elaborate on each component. \vspace{-0.5em}

\begin{figure}[t] 
\centering
\includegraphics [width=0.95\linewidth]{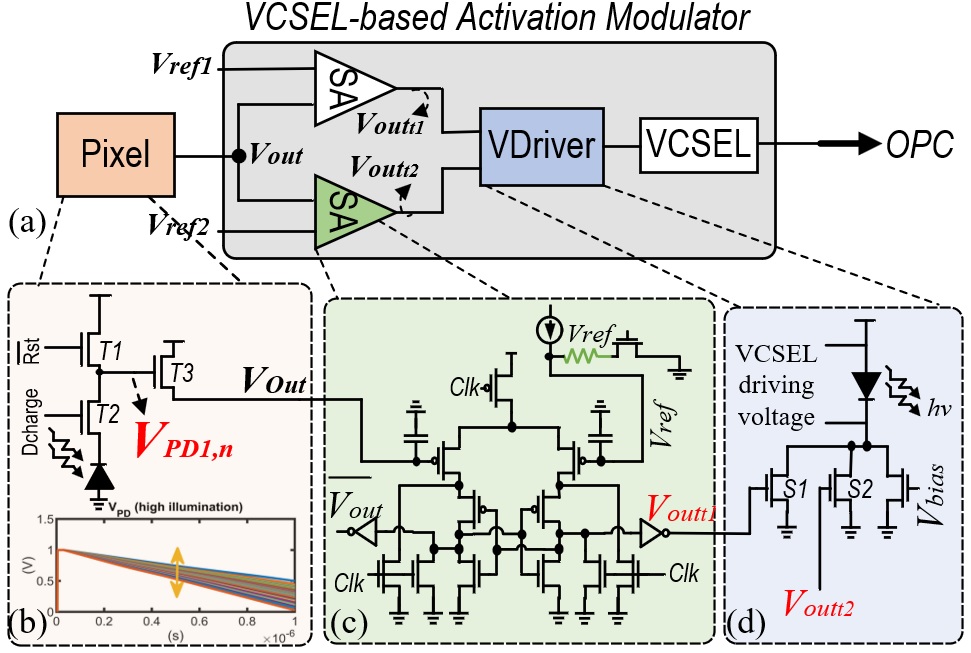}
\vspace{-1em}
\caption{(a) Proposed VCSEL-based Activation Modulator (VAM) diagram, (b) 3T pixel structure and the voltage across PD plot in high illumination, (c) Sense amplifier for thresholding, and (d) VCSEL Driver. 
}\vspace{-1.5em}
\label{VSCEL}
\end{figure}

\subsection{Microarchitectural Design }
\textbf{ADC-Less Imager.} The imager consists of pixels with a 3-transistor-1-photodiode structure as shown in Fig. \ref{VSCEL}(b) with a Photodiode (PD) as the primary sensing component, a reset transistor (T1), discharge transistor (T2), and a source–follower (T3). In the sensing mode, by initially setting Rst=`low' and Dcharge=`high', the PD connected to the T2 transistor turns into inverse polarization and fully charges the PD capacitor.  
By turning off T1, PD generates a photo-current with respect to the external light intensity which in turn leads to a voltage drop ($V_{PD}$) at the gate of T3.

\textbf{VCSEL-based Activation Modulator.} 
OISA benefits from an optimized VSCEL driver for direct activation modulation that offers power efficiency and low cost, eliminates the need for external modulators, and provides ternary input for OPC. As shown in Fig. \ref{VSCEL}(a), the VAM consists of two sense amplifiers, a modified VCSEL driver (VDriver), and the VCSEL itself. In the proposed design, the output signal from the pixel is utilized to control the biasing current of the VCSEL. By appropriately realizing one distinct reference voltage for each SA ($V_{ref}$ in Fig. \ref{VSCEL}(c)),  $V_{Out1}$, and  $V_{Out2}$ (SA outputs) generate three states, i.e., both are zero,  $V_{Out1}$ is VDD, and $V_{Out2}$ is zero, or both equal V{DD}.  $V_{Out1}$ and $V_{Out2}$ are then used to control S1 and S2 transistors which control the bising current of the VCSEL as shown in Fig. \ref{VSCEL}(d). Completely turning off the VCSEL and turning it on again to warm up imposes extra energy and delay to the design  \cite{NRZ99}. To avoid that, another biasing transistor controlled by $V_{bias}$ is added to keep the VCSEL always on. Thus, we have a non-returning-to-zero VCSEL implementation to save time and energy. The output voltages of the SAs determine the biasing current of the VCSEL and the light intensity of the generated light by the VCSEL. Accordingly, the output light intensity of the VCSEL won't be raw light, rather, it will carry ternary encoded data that corresponds to the light intensity absorbed by the pixel. In this way, our activation for the MAC operation is already modulated to the light by controlling electrical signals that run the VCSEL.

\begin{figure}[t]
\begin{center}
\begin{tabular}{ll}
\includegraphics [width=0.46\linewidth]{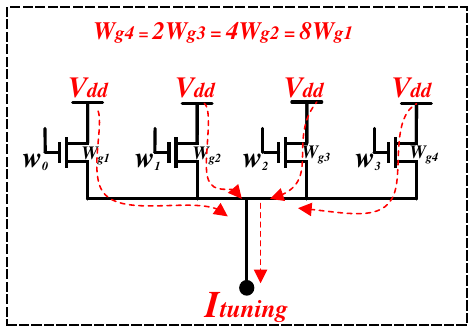} &
\includegraphics [width=0.50\linewidth]{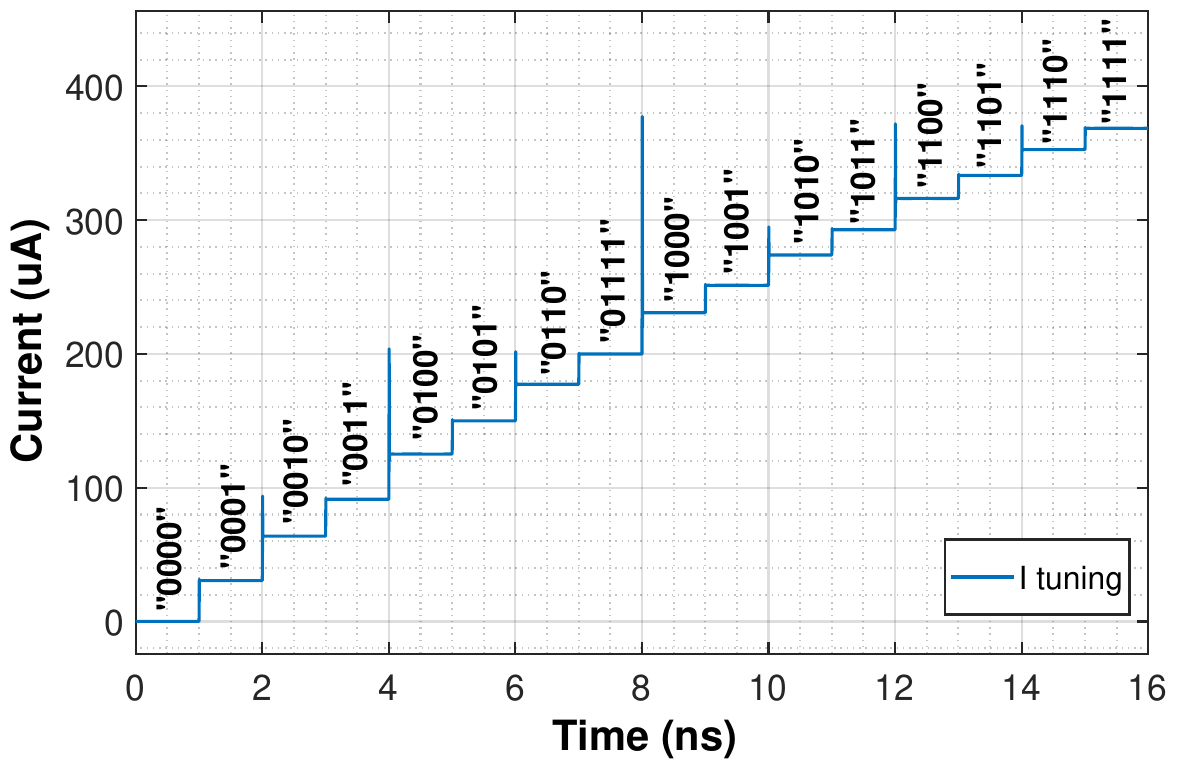}\\
\hspace{2.1 cm}  \small (a) & \hspace{2.1 cm}  \small (b)\\
 \end{tabular} \vspace{-1em}
\caption{(a) Approximate weight converter, (b) Transient simulation results.} \vspace{-2em}
\label{AWC}
\end{center}
\end{figure}

\textbf{Approximate Weight Converter.} 
In the OISA architecture, the weight parameters are initially held in on-chip kernel banks to be mapped to the MR elements in the OPC core. For this purpose, an approximate converter as shown in Fig. \ref{AWC}(a) is utilized. AWC is responsible for tuning the MRs to the desired weight values. Unlike prior optical accelerators that use area-consuming and power-hungry DACs to convert digital weight values to analog MRs' tuning signals \cite{sunny2021crosslight,zhong2023lightning,sunny2021robin},
we propose an $n$-bit approximate weight converter ($n\leq4$). As shown in Fig. \ref{AWC}(a), weight bits denoted by $w_0$ to $w_3$ are connected to the gate of T1 to T4. The key idea of AWC is to realize multi-level weighted current w.r.t. various weight values to mimic DAC behavior. Based on our circuit-level analysis, we have determined that increasing the width of transistors T1 to T4 results in a reliably enhanced current doubling effect where in the source node, all of these currents are summed. Thus, according to the spatial value of the weight bits, as depicted in Fig. \ref{AWC}(b), the AWC generates up to 16 levels ($n=4$) of current to be used for MR's tuning purposes. We elaborate on AWC latency and power consumption in Section IV.

\textbf{Optical Processing Core \& VCSEL Output Modulator.} 
The proposed OPC as shown in Fig. \ref{arch} is a non-coherent photonic computing core composed of MRs that are arrayed in arms. Each arm is comprised of 10 MRs and two waveguides used for positive and negative weights. 
As shown in Fig. \ref{arch} \encircle{1}, digital weights (W) of the DNN or MLP are converted to the analog tuning control values by the AWC unit. Utilizing these values, the weights are mapped to the MRs. This step is crucial only when OISA takes a new set of weight kernels into the processing core; once the weights are mapped, it can bypass this step. 
In \encircle{2}, activation values coming from the pixel plane are modulated to their respective wavelength using VCSELs in the VAM unit. The resultant lights whose intensity corresponds to the pixel's output values are then applied to the MR banks in the OPC. Inside the arms, each MR affects a specific wavelength of the applied light and weights the intensity of that particular wavelength. Thus, the multiplication operation of the weight values stored in the MRs and the light coming from VCSELs is conducted. 
At the end of each arm, two Balanced PhotoDiodes (BPD), shown in  Fig. \ref{arch}, perform the summation operation of both positive and negative lights resulting in an electrical output voltage that represents the result of MAC operation between the stored weights and the incoming light. In other words, BPDs convert the optical values to the electrical signals representing the sum of the dot products. Depending on the weight kernel size, these values in \encircle{3} are either summed up using extra optical summation arms or transmitted to the output of our chip directly for further processing. In the case of the MLP, the number of dot products is enormous. To reduce the complexity of the calculations, the VOM unit is added to the architecture that enables OISA to break the intensive MAC operations into smaller parts and perform the calculations. It is noteworthy that OISA uses off-chip resources to perform the non-linearity (activation function).

\textbf{MR Device Engineering.} Increasing the MR resolution ideally leads to higher weight/activation precision and accuracy. However, it demands a reduction in the Quality Factor (Q-factor), another pivotal characteristic of the MRs. The Q-factor describes the resonance's sharpness relative to the MR's central frequency. A higher Q-factor results in a sharper resonance which can make the system more sensitive to noise, as even a slight deviation in the central frequency can result in significant losses. Here, a smaller Q-factor is preferred over a large one \cite{cheng2020silicon}. Yet, achieving smaller Q-factors often entails enlarging the dimensions of the MRs, which can, in turn, introduce substantial optical crosstalk noise and energy demands for tuning. In this work, we tune and leverage the effective bit resolution of 4-bit for MRs \cite{cheng2020silicon,sunny2021crosslight} to make a balance among the above-mentioned parameters. Using Lumerical tools, we designed an MR with a radius of 5$\mu m$ and a ring waveguide width of 760 $nm$. These dimensions resulted in a relatively small Q-factor ($\approx$ 5000) which provides sufficient differentiation levels to carry out our intended multi-bit design. 
To tune MRs during weight mapping, Thermo-Optic (TO) or Electro-Optic (EO) methods are widely utilized. EO tuning is faster than TO but it can only create a slight change in MR's resonant wavelength. On the other hand, TO tuning has the capability to largely shift the MR's resonant wavelength but at a cost of larger delay and more power consumption. Similar to \cite{sunny2021crosslight}, a hybrid TO-EO tuning method is utilized to leverage both methods' advantages. \vspace{-1em}

\begin{figure}[t] 
\centering
\includegraphics [width=0.99\linewidth]{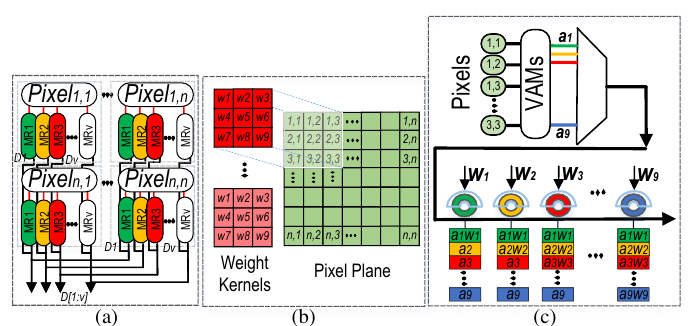}
\vspace{-2em}
\caption{(a) High-level hardware mapping to OISA, (b) Stride of a  $3\times3$ Kernel in convolution operation, (c) Implementing a stride in an arm.}
\vspace{-1.5em}
\label{mapping}
\end{figure}

\subsection{Hardware Mapping \& Bank Allocation}
\begin{figure}[t] 
\centering
\includegraphics [width=0.92\linewidth]{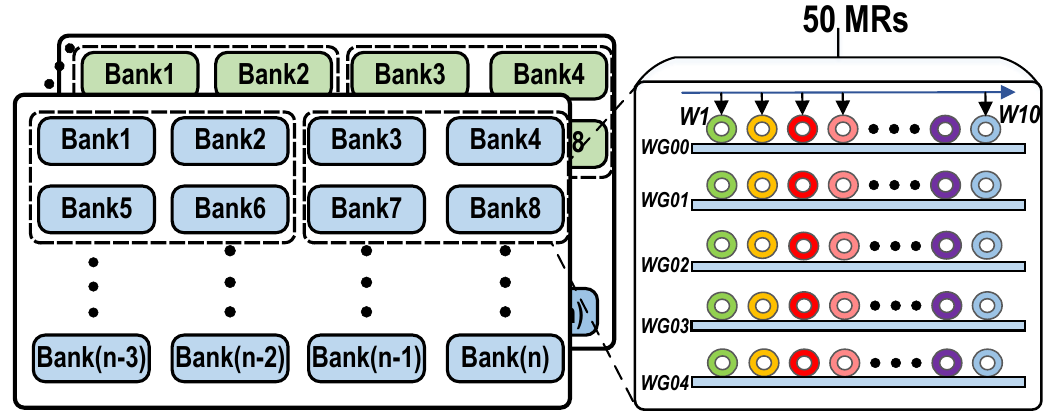}
\vspace{-1.0em}
\caption{Optical array partitioning and allocation of OISA.}
\vspace{-1em}
\label{PowerLatency}
\end{figure}

Aiming to process the first layer of DNNs, OISA is equipped with a correlated hardware mapping method to balance the workload and increase the throughput. A high-level hardware mapping is shown in Fig. \ref{mapping}(a), where the inputs from the pixel plane are connected to their respective weight values that are mapped to the MRs. A 3$\times$3 kernel stride over a pixel plane is shown in Fig. \ref{mapping}(b). To localize the kernel stride computation in one arm, OISA supports a kernel size of 3$\times$3 as seen by several well-known
DNN models. Fig. \ref{mapping}(c) shows the detailed implementation of a stride where weight values related to a 3$\times$3 kernel are mapped on the MRs in an arm. Input activation values then are modulated on a light with a wavelength according to the resonance wavelength of their corresponding weight. These input-modulated lights are then multiplexed to a single light and are passed through the arm. MRs affect the intensity of the light passing the arm and weight the specific wavelength providing multiplication between 9 activation values and 9 weight values. Later a BPD at the end of the arm will provide the optical summation of multiplication results to conclude the MAC operation.
We develop OISA with 10 MRs in each arm that enables us to perform 9$\times$9 (3$\times$3 kernel values by 9 activation values) MAC operations. However, in order to enhance the core's efficiency in handling 5$\times$5 and 7$\times$7 kernels (25 and 49 weight values), we partition the cores in the OPC hierarchy. As shown in Fig. \ref{PowerLatency}, in our core, each bank comprises 5 arms, each equipped with 10 MRs, resulting in a total of 50 MRs per bank. With 80 banks in OPC, OISA consists of 4000 MRs in total. Banks are grouped in the 4 columns. Thus, each row has 40 MRs, and 40 AWC units are assigned to map the weights to MRs. It is worth pointing out that to completely map all the weight values to the OPC, 100 iterations are required. In the case of 3$\times$3 kernels, the MAC result of each arm will represent a stride of the convolution operation and can be directly transferred to the output. Supporting 5$\times$5 and 7$\times$7 weight kernels, the output of each bank will be further processed in the VOM unit to obtain the final MAC results. According to this configuration, the total number of MAC operations that can be processed in one cycle ($N_{\frac{m}{cycle}}$) can be formulated as $f\times(n K^2)$. Where $f$ is the number of banks, $K\in\{3,5,7\}$ is the kernel size.
We consider $n=5$ when $K=3$, as 5 kernels with the size of 3$\times$3 can be mapped to each bank. Else $n=1$ as only one 5$\times$5 or 7$\times$7 can be mapped to each bank. Thus, for $K=3, 5, 7$, in each cycle, OISA conducts 3600, 2000, and, 3920 MAC operations.
The total required cycles for performing convolution operation depend on the number of weight kernels and their sizes. \vspace{-0.5em}

\section{Performance Evaluation}
\begin{figure}[t] 
\centering
\includegraphics [width=0.82\linewidth,height=4.7cm]{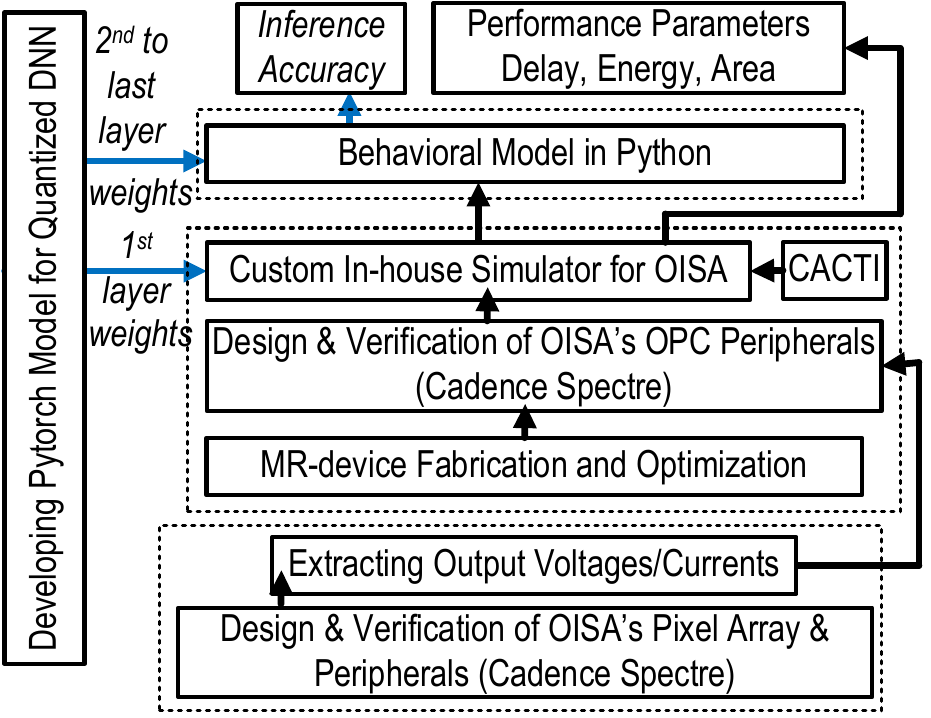}
\vspace{-1em}
\caption{Proposed bottom-up evaluation framework.}
\vspace{-2em}
\label{framework}
\end{figure}

\vspace{-0.2em}
\textbf{Framework.} The evaluation framework is created through a bottom-up methodology as shown in Fig. \ref{framework}. At the device level, we fabricated and optimized the MR device and extracted the circuit parameters to co-simulate with interface CMOS circuits in Cadence Spectre and SPICE.
At the circuit level, we first implemented the OISA pixel's array and peripheral circuitry using the 45nm NCSU Product Development Kit (PDK) library \cite{NCSU_PDK} in Cadence and extracted the output voltages and currents. We then developed all OISA's components except the kernel banks (implemented in Cacti \cite{thoziyoor2008cacti}) in Cadence Spectre.
The DNN weight parameters associated with the 1$^{st}$ layer need to be quantized and mapped into the OPC, while the remaining layers are processed with the off-chip processors. We trained a PyTorch model w.r.t. the under-test datasets and extracted the 1$^{st}$-layer weights. OISA's MR elements are then adjusted with these weights. For computation purposes, we developed a custom in-house simulator for OISA. This simulator computes the overall latency and power consumption required for the network execution. It also offers flexibility in array configuration and peripheral design selection. The results are captured after processing the 1$^{st}$ convolution layer. We then developed a Python-based behavioral DNN model to utilize the computation outcomes and processes the 2$^{nd}$-to-last layer to calculate inference accuracy.

\begin{figure}[b]  \vspace{-1em}
\centering
\includegraphics [width=0.79\linewidth,height=6cm]{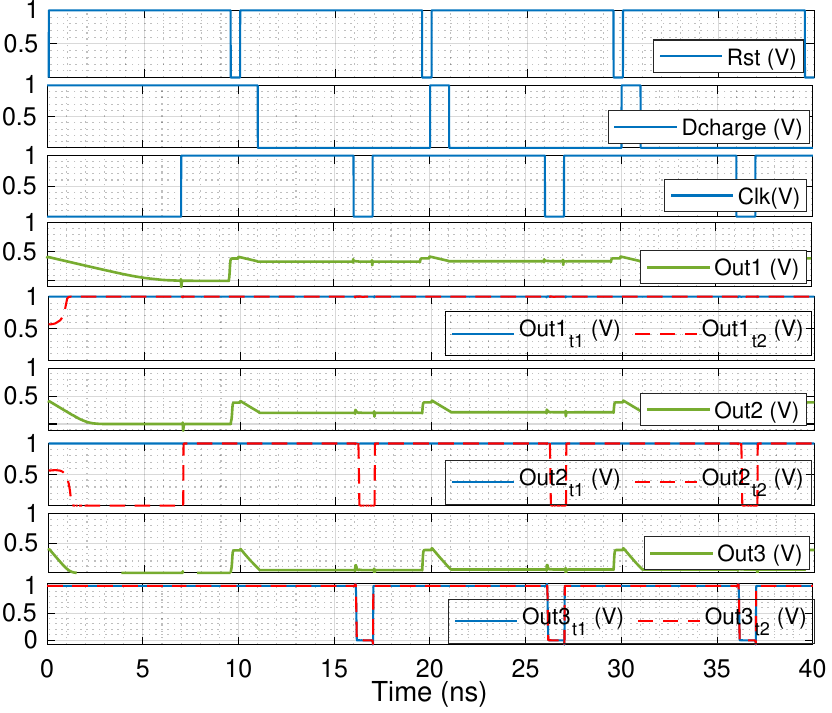}
\vspace{-1.5em}
\caption{Transient simulation results of VAM's thresholding to drive VCSEL.}
\vspace{-1em}
\label{waveform}
\end{figure}

\textbf{Functionality.} 
Figure \ref{waveform} shows the transient simulation waveforms of VAM's thresholding to drive VCSEL depicted in Fig. \ref{VSCEL}(a). 
In this figure, the $t_1$ and $t_2$ represent two outputs corresponding to three distinct pixels denoted as $Out1$, $Out2$, and $Out3$. Each pixel exhibits a unique voltage implying its absorption of varying light intensities. Specifically, the voltage levels for $Out1-3$ are ascertained whenever the Clk signal is low, observable within the time frame of 16 to 17 ns. Within this interval, the voltage for $Out1$ surpasses both sense amplifier thresholds, resulting in both $t_1$ and $t_2$ being set to 1. Conversely, the voltage for $Out2$ resides between 0.16V and 0.32V, leading to $t_1$ equating to 1 and $t_2$ being 0. As for $Out3$, its voltage is less than 0.16V, thereby setting both $t_1$ and $t_2$ to 0.

\color{black}

\begin{figure}[t] 
\centering
\includegraphics [width=0.96\linewidth]{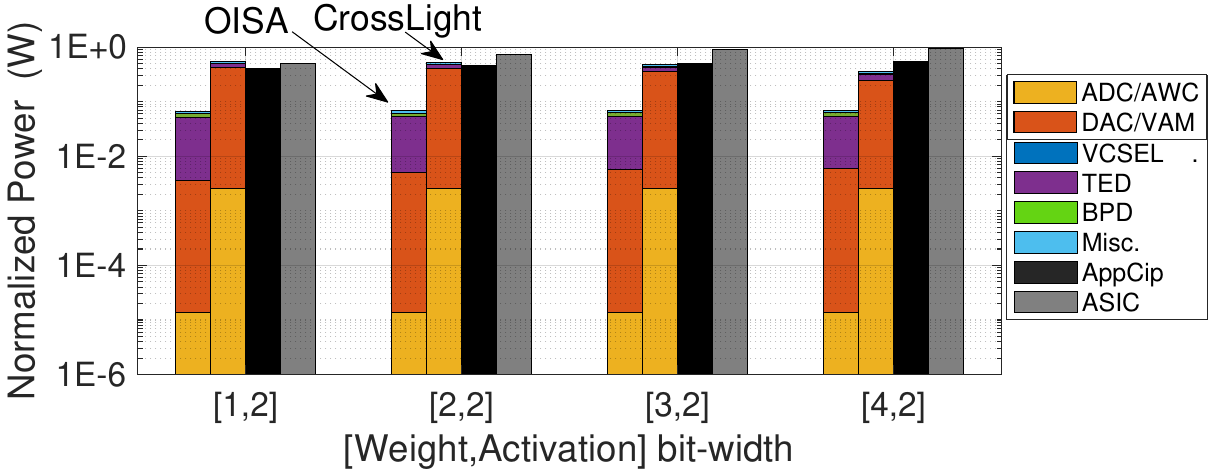}
\vspace{-1.0em}
\caption{Normalized log-scaled power consumption of various accelerators. From left to right: OISA, Crosslight, AppCip, and ASIC design.}
\vspace{-1.5em}
\label{Power}
\end{figure}

\textbf{Power Consumption \& Performance.} 
Here, we compare the OISA with three DNN accelerators based on PIS and ASIC as follows assuming a 1-to-4-bit width for the weight parameter. For evaluation, we assume that all platforms process the 1$^{st}$ layer of the ResNet18 model.
\ul{\textit{\textbf{Optical PIS}:}} We designed a Crosslight-like \cite{sunny2021crosslight} platform with 80 banks, each consisting of 5 arms with 10 MRs. For a fair evaluation, we developed the design from scratch using the proposed evaluation framework and the in-house simulator to extract numbers. Note that the Crosslight uses separate banks for weight and activation.
\ul{\textit{\textbf{Electronic PIS}:}} We developed an AppCip-like \cite{tabrizchi2023appcip} accelerator with non-volatile memory in HSPICE and NVSIM \cite{dong2012nvsim} from scratch and extracted the performance parameters.
\ul{\textit{\textbf{ASIC}:}} We developed a DaDianNao-like \cite{chen2014dadiannao} accelerator with 8$\times$8 tile version connected to a conventional 128$\times$128 image sensor. We synthesized the designs with the Design Compiler under the 45 nm process node. The eDRAM and SRAM performance was estimated using CACTI \cite{thoziyoor2008cacti}.
Figure \ref{Power} shows the normalized power consumption of the under-test platforms in various bit-width configurations. From this figure, we observe the superiority of OISA over various under-test platforms where it achieves 8.3$\times$, 7.9$\times$, and 18.4$\times$ reduction in power consumption on average compared with Crosslight \cite{sunny2021crosslight}, AppCip \cite{tabrizchi2023appcip}, and ASIC platforms, respectively. We report the breakdown of power consumption for OISA and Crosslight platforms as well, where we observe a remarkable reduction in power consumption mainly due to ADC/DAC elimination in OISA as compared with Crosslight. 
As for execution time, considering that the activation and weight values are already mapped to the core, in the OISA, and Crosslight-like designs, the utilized VCSEL and BPD technologies have critical effects on the execution time.  To have a fair comparison same VCSEL \cite{flipchip} and BPD \cite{sunny2021robin} technologies have been used for OISA and corsslight-like designs. The total execution time for performing one architecture-wide MAC operation is 55.8 ps which results in 7.1 TOp/s. However, the most important difference between these two designs is that in the OISA, all of the MRs in OPC are allocated to weight values, while in the Crosslight-like design, half of the MRs are considered to be mapped by activation which cuts the total number of operations to half.  

\begin{table*}[t]
\centering
\caption{Performance comparison of various 
PIS/PNS/PIP units. Note: OISA is the only hybrid in-sensor accelerator.} \vspace{-1em}
\scalebox{0.75}{
\begin{tabular}{ccccccccccc}
\hline
\rowcolor[HTML]{C0C0C0} 
\textbf{Designs}                       & \textbf{\begin{tabular}[c]{@{}c@{}}Technology\\ (nm)\end{tabular}} & \textbf{Purpose}                                                            & \textbf{Comput. Scheme} & \textbf{Memory} & \textbf{NVM} & \textbf{\begin{tabular}[c]{@{}c@{}}Pixel Size\\ ($\mu m^2$)\end{tabular}} & \textbf{Array Size} & \textbf{\begin{tabular}[c]{@{}c@{}}Frame Rate\\ (frame/s)\end{tabular}} & \textbf{\begin{tabular}[c]{@{}c@{}}Power\\ (mW)\end{tabular}}               & \textbf{\begin{tabular}[c]{@{}c@{}}Efficiency\\ (TOp/s/W)\end{tabular}} \\ \hline
\textbf{\cite{park20147}}              & 180                                                                & 2D optic flow est.                                                          & row-wise                & Yes             & No           & 28.8$\times$28.8                                                          & 64$\times$64        & 30                                                                      & 0.029                                                                       & 0.0041                                                                  \\ \hline
\textbf{\cite{hsu20200}}               & 180                                                                & \begin{tabular}[c]{@{}c@{}}edge*/blur/sharpen/\\ 1$^{st}$-layer CNN\end{tabular} & row-wise                & No              & No           & 7.6$\times$7.6                                                            & 128$\times$128      & 480                                                                     & \begin{tabular}[c]{@{}c@{}}sensing: 77 \\ processing: 91\end{tabular}       & 0.777                                                                   \\ \hline
\textbf{\cite{yamazaki20174}}          & 60/90                                                              & STP$^\dagger$                                                               & row-wise                & Yes             & No           & 3.5$\times$3.5                                                            & 1296$\times$976     & 1000                                                                    & \begin{tabular}[c]{@{}c@{}}sensing: 230 \\ processing:363\end{tabular}      & 0.386                                                                   \\ \hline
\textbf{\cite{xu2020macsen}}           & 180                                                                & 1$^{st}$-layer BNN                                                               & entire-array            & Yes             & No           & 110$\times$110                                                            & 32$\times$32        & 1000                                                                    & 0.0121                                                                      & 1.32                                                                    \\ \hline
\textbf{\cite{carey2013100}}           & 180                                                                & edge*/TMF$^\ddagger$                                                        & row-wise                & Yes             & No           & 32.6$\times$32.6                                                          & 256$\times$256      & 100,000                                                                 & 1230                                                                        & 0.535                                                                   \\ \hline
\textbf{\cite{angizi2023pisa}}         & 65                                                                 & 1$^{st}$-layer BNN                                                               & entire-array            & Yes             & Yes          & 55$\times$55                                                              & 128$\times$128      & 1000                                                                    & \begin{tabular}[c]{@{}c@{}}sensing: 0.025\\ processing: 0.0088\end{tabular} & 1.745                                                                   \\ \hline
\textbf{\cite{xu2021senputing}}        & 180                                                                & 1$^{st}$-layer BNN                                                               & entire-array            & Yes             & No           & 35$\times$35                                                              & 32$\times$32        & 156                                                                     & 0.00014 - 0.00053                                                           & 9.4-34.6                                                                \\ \hline
\textbf{\cite{lefebvre20217}}          & 65                                                                 & 2 - 64 Conv/ROI**                                                           & row-wise                & No              & No           & 9$\times$9                                                                & 160$\times$128      & 96 - 1072                                                               & 0.042 - 0.206                                                               & 0.15 - 3.64                                                             \\ \hline
\textbf{\cite{song2022reconfigurable}} & 180                                                                & 1$^{st}$-layer CNN                                                               & entire-array            & No              & No           & 10$\times$10                                                              & 128$\times$128      & 3840                                                                    & 0.45 - 1.83                                                                 & 1.41 - 3.37                                                             \\ \hline
\textbf{\cite{tabrizchi2023appcip}}                        & 45                                                                 & 1$^{st}$-layer CNN                                                               & entire-array            & Yes             & Yes          & 38$\times$38                                                              & 32$\times$32        & 3000                                                                    & 0.00096 - 0.0028                                                                     & 1.37 - 4.12                                                                    \\ \hline
\textbf{OISA}                        &         65                                                          & 1$^{st}$-layer CNN                                                               & entire-array            & Yes             & No          &  4.5$\times$4.5                                                             &   128$\times$128      &    1000                                                                 &   0.00012-0.00034                                                                   &  6.68                                                                   \\ \hline
\end{tabular}}
\label{comp}

$^*$ Edge extraction.
$^\dagger$Spatial Temporal Processing.
$^\ddagger$Thresholding Median Filter. 
$^{**}$Region Of Interest.
\vspace{-2em}
\end{table*}


Table \ref{comp} presents a comparison of the structural and performance characteristics of selective PIS implementations in the electronic domain and OISA. Since these implementations are tailored for specific domains, we conducted a fair assessment by estimating and normalizing the power consumption considering a scenario where all PIS units process the first layer of a CNN. The OISA reaches the frame rate of 1000 and the efficiency of $\sim$6.68 TOp/s/W as one of the most efficient implementations. This comes from the massively-parallel OPC banks and eliminating ADC/DAC for coarse-grained inference.  As for the area, according to the MR's dimensions mentioned in section III, and our architecture configurations, the total area of the OISA is 1.92 $mm^2$. The simulation results reported in Table \ref{comp} demonstrate no modification on the pixel array. Due to our lack of access to the configurations of other layouts, it is challenging to establish a fair comparison with respect to the total area overheads.


\textbf{Accuracy.} We conduct experiments on OISA considering various $[$Weight: Activation$]$ configurations with several datasets, including MNIST evaluated on LeNet, SVHN on ResNet18, CIFAR-10 on ResNet18, and CIFAR-100 on VGG16 compared with a software baseline, FBNA \cite{FBNA}, AppCiP \cite{tabrizchi2023appcip}, and PISA \cite{angizi2023pisa} as recent low bit-with accelerators. The comparison results of classification accuracy are summarized in Table \ref{accuracyT}. We find that 1) OISA shows an acceptable accuracy while providing significant power-delay reduction as discussed earlier compared with other platforms; 2) Generally, our experiments show that weights and inputs are progressively more sensitive to bit-width changes. However, a higher weight bit-width does not necessarily result in a higher accuracy as indicated in the OISA $[$4:2$]$ configuration. This comes from the fact that AWC may not reliably provide distinct current levels when the number of bits increases; and 3) The accuracy drop of OISA is mainly because of the ADC-DAC-less nature of it that allows processing the 1$^{st}$ convolutional layer with 1 to 4 bits approximated with a converter. 
\vspace{-1em}

\begin{table}[h]
\centering
\caption{OISA's Accuracy (\%) on various datasets.}
\vspace{-1em}
\scalebox{0.75}{
\begin{tabular}{cccccc}
\rowcolor[HTML]{C0C0C0} 
\hline
 \textbf{Configuration} &  \textbf{MNIST}  &   \textbf{SVHN}    &   \textbf{CIFAR-10} &  \textbf{CIFAR-100}   \\ \hline
baseline   &    99.6   &     97.5  &     91.37    &  78.4 \\
FBNA \cite{FBNA}   &    --   &  96.9     &   88.61      & \textcolor{black}{71.5}  \\
AppCiP \cite{tabrizchi2023appcip}   &    --   &  96.4     &   89.51      & \textcolor{black}{--}  \\
PISA \cite{angizi2023pisa}   &  95.12   &   90.35  &  79.80 &  \textcolor{black}{61.6}  \\ 
OISA $[$4:2$]$   &  95.21  &  91.74   &  81.23 & 61.38 \\ 
OISA $[$3:2$]$   &  96.18  &  94.36   &  84.45 & 66.89 \\ 
OISA $[$2:2$]$  &   96.25  &  93.20   & 83.85  & 66.94  \\ 
OISA $[$1:2$]$   & 95.75   &  93.16   &  83.64 & 66.06  \\ \hline
\end{tabular}}
\label{accuracyT}
\end{table}

\vspace{-0.5em}
\section{Conclusion}
In this work, we presented OISA as a high-performance and energy-efficient optical in-sensor accelerator architecture. OISA benefits from an innovative design and hardware mapping method to remarkably reduce the power consumption of data conversion, transmission, and processing in the conventional cloud-centric architecture as well as recent edge accelerators. Our results on various image data-sets show acceptable accuracy while OISA achieves 6.68 TOp/s/W efficiency.

\section{Acknowledgment}
\small This work is supported in part by the National Science Foundation (NSF) under grant numbers 2046226, 2006788, 2216772, 2216773, and 2247156.

\bibliographystyle{IEEEtran} \vspace{-0.7em}
\bibliography{IEEEabrv,./main}\vspace{-2em}
\vspace{-2em}
\end{document}